\documentclass[jcp,aip,reprint,showpacs,floatfix,groupedaddress]{revtex4-1}
\usepackage{graphicx}
\usepackage{dcolumn}
\usepackage{amssymb}
\usepackage{bm}

\def\s#1{_{\rm #1} }
\def\sp#1{^{\rm #1} }
\def\vep{\varepsilon}
\def\bea{\begin{eqnarray}}
\def\eea{\end{eqnarray}}
\def \be{\begin{equation}}
\def \ee{\end{equation}}

\begin{document}
\title{Photoferroelectric solar to electrical conversion}
\author{Milo\v{s} Kne\v{z}evi\'{c}}
\author{Mark Warner}
\email{mw141@cam.ac.uk}
\affiliation{Cavendish Laboratory, University of Cambridge, JJ Thomson Avenue,
Cambridge CB3 0HE, United Kingdom}
\date{\today}

\begin{abstract}
We propose a charge pump which converts solar energy into DC
electricity. It is based on cyclic changes in the spontaneous electric
polarization of a photoferroelectric material, which allows a transfer of charge from a low
to a high voltage. To estimate the power efficiency we use a photoferroelectric liquid
crystal as the working substance. For a specific choice of material, an efficiency
of $2\%$ is obtained.
\end{abstract}

\pacs{61.30.Gd, 77.84.Nh, 77.80.B-, 84.60.Jt, 85.50.-n, 88.40.H-}

\maketitle


Photoferroelectrics are materials in which their ferroelectricity can be
affected by exposure to light. We propose a device that works as a charge
pump based on the photoferroelectric effect. Varying light intensity cycles a
photoferroelectric working material through various electric polarization states and the
associated charge in the external circuit is pumped from low to high potential.

\begin{figure}
  \includegraphics[width=7.8cm]{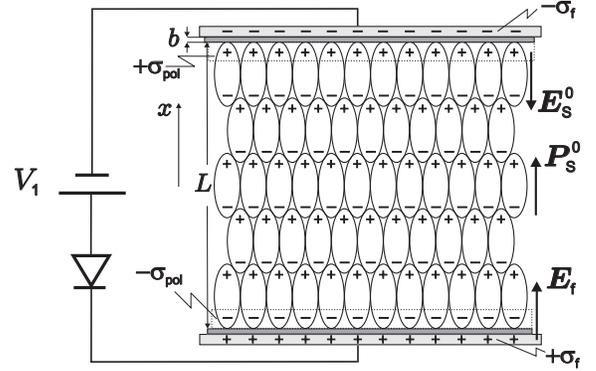}
  \caption{A slab of ferroelectric of thickness $L$ with two insulating layers
of thickness $b$ under the bounding electrodes.  A polarization surface charge
layer $\sigma\s{pol}$ associated with truncating the spontaneous polarization
$\bm{P}\s{s}^0$ generates an internal field $\bm{E}\s{s}^0$. A free surface counter
charge layer $\sigma\s{f}$ partially neutralizes the polarization charges to give a
resultant electric field commensurate with an applied voltage $V_1$. The electric
field $\bm{E}\s{f}$ of charges $\sigma\s{f}$ is also indicated.}
  \label{fig1}
\end{figure}

A body with a uniform (ferroelectric) polarization $\bm{P}\s{s}^0$
has zero polarization charge density in the bulk, $\rho\s{pol} = - \bm{\nabla}\cdot
\bm{P}\s{s}^0 = 0$, while the surface charge density associated with
the change of polarization from its finite bulk value to zero on traversing the surface
cutting $\bm{P}\s{s}^0$ is given by $\sigma\s{pol} = P\s{s}^0$.
Figure~\ref{fig1} shows this $\sigma\s{pol}$ associated with truncating
$\bm{P}\s{s}^0$.  Such a layer of charge $\sigma\s{pol}$ creates an internal electric field
$\bm{E}\s{s}^0= - \bm{P}\s{s}^0/\vep_0 \vep$ in the material in this slab geometry,
where $\vep$ is the relative dielectric constant of the material.
In the presence of the electrodes on the surfaces of the slab that are shorted
together ($V_1 = 0$ in Fig.~\ref{fig1}), and the absence of the insulating layers,
external free charges of surface density $\sigma\s{f} = - \sigma_{pol}$
flow to neutralize $\sigma\s{pol}$. These free charges vitiate the field
$\bm{E}\s{s}^0$ between the plates by creating an opposite field, $\bm{E}\s{f}$,
that is in the direction of $\bm{P}\s{s}^0$. See Fig.~\ref{fig1} for counter charge
layers $\pm \sigma\s{f}$ on the bounding electrodes. In fact, they have been further
adjusted from $\mp \sigma\s{pol}$ in order to increase $\bm{E}\s{f}$ so
that the net field matches the actual applied potential $V_1 \ne 0$ instead.
It is such free, external balancing charges we aim to optically pump.

Considering the two electric fields, the voltage $V_1$ can be expressed in the form
\be
V_1 = \int {\rm d}x \left( E\s{f} - E\s{s}^0\right) = \sigma\s{f} \left ( \frac{2b}{\vep_0 \vep_1} + \frac{L}{\vep_0 \vep} \right ) - \frac{P\s{s}^0L}{\vep_0 \vep},
\label{voltdark}
\ee
where $L$ is the thickness of the photoferroelectric sample itself,
$b$ is the thickness of insulating layers inserted between the sample and the electrodes to prevent charge flow,
and $\vep_1$ is the layers' relative dielectric constant.
If $P\s{s}^0$ were now diminished, then the internally-generated field $E\s{s}^0$ would also be diminished.
Now the field associated with the free, formerly partially neutralizing charges, if they cannot
dissipate, generates an increased potential $V\s{on}>V_1$ between the electrodes. This increased potential $V\s{on}$
allows charge pumping.

The spontaneous polarization of a photoferroelectric usually decreases~\cite{ikeda:93n,giesselmann:03,giesselmann:04} under light
irradiation as photons are absorbed and modify the polar ordering. Since light
is absorbed the beam is necessarily attenuated and the polarization profile can
no longer be uniform (except when light is powerful enough that $P\s{s}(x) = 0$
at all depths $x$). Then $\bm{\nabla} \cdot \bm{P}\s{s}$ is non-zero also in the bulk,
and polarization charges will emerge inside the sample as well as at the surface.
Consequently a non-uniform electric field $\bm{E}\s{s}(x,t) = -\bm{P}\s{s}(x,t)/\vep_0 \vep$
appears in the sample, where $\bm{P}\s{s}(x,t)$ is the electric polarization at the point $x$
and the illumination time $t$; note that $E\s{s}(x,t) \leq E\s{s}^0$.
As the diode prevents backflow of the charge to the battery (see Fig.~\ref{fig1}),
the resulting voltage $V\s{on}(t)$ between the electrodes will be greater than $V_1$:
\be
V\s{on}(t) = \sigma\s{f} \left ( \frac{2b}{\vep_0 \vep_1} + \frac{L}{\vep_0 \vep} \right )
- \frac{1}{\vep_0 \vep}\int_{-L/2}^{L/2} P\s{s}(x,t) {\rm d} x,
\label{von}
\ee
where the distance $x$ is measured from the center of the sample (Fig.~\ref{fig1}).
Writing $\overline{P\s{s}}(t) = \int_{-L/2}^{L/2} P\s{s}(x,t) {\rm d} x/L$ for
the spatial average of the polarization profile after time $t$, and replacing $\sigma\s{f}$
with the aid of Eqn.~(\ref{voltdark}), we get
\be
V\s{on}(t) = V_1 + \frac{L}{\varepsilon_0 \varepsilon} \left ( P\s{s}^0 - \overline{P\s{s}}(t) \right ).
\label{vonc}
\ee

The described mechanism can be exploited to pump charge from a low to high
voltage battery. Many different materials exhibit photoferroelectric behavior, and
might be potential candidates for use in our device. We therefore
explore the suitability of these materials for a charge pump, optimize their performance by specifying
an appropriate electric circuit, and calculate the efficiency of such pumps.
Our general analysis is applicable to all photoferroelectric materials.
Only our final estimate of the efficiency of the pump will be demonstrated for
particular substances, photoferroelectric liquid crystals.

\begin{figure}
  \includegraphics[width=7.8cm]{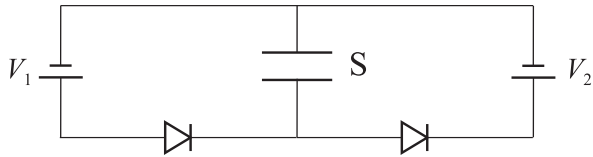}
  \caption{Charge pump diagram with S being the photoferroelectric sample.
 Charge is pumped from the battery $V_1$ to higher voltage $V_2$.}
  \label{fig2}
\end{figure}

The circuit diagram of the charge pump is shown in Fig.~\ref{fig2}. The central
component of this circuit is the photoferroelectric sample S placed between transparent conducting
electrodes. The arrangement of diodes is such that the charge can flow from the voltage $V_1$ to
a higher voltage $V_2$ only. We shall describe how the electric polarization varies during a cycle.

The photoferroelectric sample S in the dark state is connected to the voltage $V_1$.
When light is shone on the sample, the electric polarization decreases and the voltage
across the electrodes is increased above the value $V_1$,
with charge on the electrodes being fixed (the left diode prevents backflow of
charge). As the electric polarization further decreases, the voltage across the electrodes
increases until it reaches the value $V_2$.
Then the charge starts to flow through the right diode, which leads to its
being pumped into battery $V_2$ at constant voltage (Fig.~\ref{fig2}). If the
connection of the sample with the battery $V_2$ were absent, the voltage across the sample
would rise to a value we simply denote by $V\s{on}$ corresponding to $V\s{on}(t)$ with
$t=t\s{on}$, the duration of the illumination phase of the cycle.

The charge transferred to the battery $V_2$ is equal to $\Delta q = C (V\s{on} - V_2)$, where
$C$ is the total capacitance of the sample.
There are two insulating layers of capacitance $C\s{b}$ and a liquid crystal of
capacitance $C\s{L}$ connected in series. The equivalent capacitance is given
by $1/C = 2/C\s{b} + 1/C\s{L} = (2b/\vep_1 + L/\vep)/(\vep_0 A)$, where $A$ is the area of the electrode.
Taking into account that $b \ll L$, and typically $\vep \sim \vep_1$, one
gets a simple expression $C= \vep_0 \vep A/L$.

Upon switching off the light, the spontaneous polarization will evolve to a new
state. Here, we shall suppose that it recovers its initial dark state value; this, in
particular, holds for a large class of photoferroelectric liquid crystals.
As the polarization increases in time, for a constant value of the
charge on the electrodes, the voltage across the sample starts to decrease from
its value $V_2$. When the voltage drops to $V_1$, new charges will start to flow from the left battery
to maintain the voltage $V_1$ across the electrodes. Finally, the charge on the electrodes recovers its
value at the beginning of the cycle. This cycle is reminiscent of an optical charge pump based on a
capacitance that is optically variable.~\cite{hiscock:11} In practice, a self-priming circuit~\cite{mckay:10}
would avoid exhausting battery $V_1$.

The same total charge $\Delta q $ pumped into the battery $V_2$ is
supplied by $V_1$ to the electrodes during recovery process. Thus, the work
output $W$ of this charge pump is $W = \Delta q (V_2 - V_1)= \Delta q \Delta V$, which gives
\be
W = \vep_0 \vep \frac{A}{L} \Delta V \left [ \frac{L}{\vep_0 \vep} \left (
P\s{s}^0 - \overline{P\s{s}}(t\s{on}) \right ) - \Delta V \right ].
\label{work}
\ee
We can maximize this output with respect to the voltage difference
$\Delta V$. Using the condition ${\rm d}W/{\rm d} \Delta V = 0$, we get the
optimal choice of voltage difference
\be
(V_2 - V_1)\sp{opt} = \frac{L}{2 \vep_0 \vep} \left ( P\s{s}^0 -
\overline{P\s{s}}(t\s{on}) \right ) = \frac{V\s{on} - V_1}{2}.
\label{ovolt}
\ee
Since to pump charge we require the open circuit developed voltage to exceed
that of the upper battery, that is $V_2 < V\s{on}$, or equivalently $V_2 - V_1
< V\s{on} - V_1$, we see that the above choice of voltage difference is
allowed. Thus, for given voltages $V_1$ and $V\s{on}$, the optimal $V_2$ is
\be
V_2\sp{opt} = \frac{V_1 + V\s{on}}{2}.
\label{optimal}
\ee
The maximum output $W\s{m}$ is then
\be
W\s{m} = \frac{AL}{4 \vep_0 \vep} \left ( P\s{s}^0 - \overline{P\s{s}}(t\s{on})
\right )^2.
\label{owork}
\ee
Output is therefore higher if the dark state polarization $P\s{s}^0$ and the average value
of the polarization profile $\overline{P\s{s}}(t\s{on})$ in the illuminated state differ more, that is if as
much polarization as possible is lost. The work delivered depends on the volume $AL$ of the working material.

If $I\s{sun}$ is the incident solar flux, and $U\s{in}$ is the energy input per unit volume of the sample
required to transform the polarization state of the working material, then $t\s{on}$ is determined by
energy balance $I\s{sun} A t\s{on} = U\s{in} A L$. The period of a cycle $\tau = t\s{on} + t\s{off}
= U\s{in}L/I\s{sun} + t\s{off}$ then follows. Here, $t\s{off}$ is the characteristic time for relaxation of polarization
to its dark state value. The power efficiency is then $\eta = W\s{m}/(\tau I\s{sun} A)$.

The power efficiency of this cycle is significantly reduced, due to the fact that no work is
done during the time $t\s{off}$. To avoid such kind of difficulties one can use an array of
photoferroelectric samples.
They could be sequentially illuminated and then allowed to relax back to the dark state,
either via a movable focusing mirror, or by rotating an array of samples.
It is clear, however, that the described setting will reduce to some extent the desired enhancement of power efficiency.
As typically $t\s{on} < t\s{off}$, one should use an array of at least $m=t\s{off}/t\s{on}$ samples.
For such a device the power efficiency $\eta$ is given by the above formula, but in which the cycle
time $\tau$ is equal to $t\s{on}$, that is
\be
\eta = \frac{W\s{m}}{U\s{in}AL} = \frac{(P\s{s}^0 - \overline{P\s{s}}(t\s{on}))^2}{4\vep_0 \vep U\s{in}}.
\label{eta}
\ee
To obtain a numerical estimate of the power efficiency (\ref{eta}) we confine ourselves to photoferroelectric
liquid crystals, the main properties of which we sketch.

\begin{figure}
  \includegraphics[width=7.2cm]{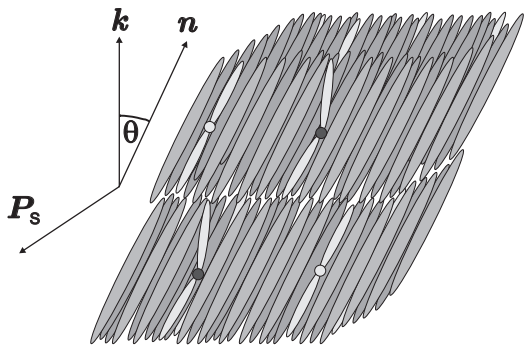}
  \caption{A smectic C* liquid host. Guest dye molecules with a photoactive center indicated by a circle
  are also shown, some of them in the linear (\textit{trans}) state, and others in the bent (\textit{cis})
  excited state.}
  \label{fig3}
\end{figure}

Liquid crystals are anisotropic fluids which usually consist of rod-like molecules. The
liquid-crystalline states are observed at temperatures between the solid state and the
ordinary isotropic liquid state.
In the simplest nematic phase long-range translational order is absent, but the
long molecular axes are orientationally ordered, on average about a unit axial vector $\bm{n}$, the director.
In addition to nematic orientational order, the molecules in smectic phases are arranged into
layers. In the smectic A phase (SmA), the molecules are organized as two-dimensional anisotropic
liquids within the layers, with the nematic director $\bm{n}$ being parallel to the normal $\bm{k}$
of the layers. In the smectic C phase (SmC), $\bm{n}$ is no longer parallel to $\bm{k}$, but is
tilted by an angle $\theta$ with respect to $\bm{k}$ (Fig.~\ref{fig3}). The tilt angle $\theta$ is a function
of temperature $T$. At a given $T$, the director $\bm{n}$ is located on the surface of a cone
centered around $\bm{k}$ with opening angle $\theta(T)$.

A SmC liquid crystal composed of chiral molecules is referred to as a SmC* (Fig.~\ref{fig3}).
As first pointed out by Meyer~\cite{meyer:75}, the emergence of a (local) spontaneous electric polarization
$\bm{P}\s{s}$ in SmC* is allowed along $\bm{k} \wedge \bm{n}$ since the plane of $\bm{k}$ and
$\bm{n}$ is not a mirror plane. The spontaneous polarization is a function of the tilt angle,~\cite{lagerwallbook:99}
and $P\s{s} \propto \theta$ for small $\theta$.
When the temperature is increased and crosses a threshold value, the SmC* phase most often undergoes a second-order phase
transition to the non-polar SmA* phase.

The spontaneous polarization of a SmC* phase doped with photosensitive molecules (for example, an azo or thioindigo dye) can be
changed by light irradiation;~\cite{ikeda:93n,giesselmann:03} pure systems of dyes have also been used.~\cite{giesselmann:04}
Some dye molecules can make transitions from the linear (\textit{trans}) ground state to the excited bent-shaped
(\textit{cis}) state by absorbing a photon. The \textit{cis} state isomers can disrupt the SmC* ordering,
which leads to a change of $P\s{s}$, whence the name photoferroelectric effect.
If the interaction with light is sufficiently strong it can completely destroy
the SmC* phase, causing a transition to the non-polar SmA* phase.
Dye molecules in the excited \textit{cis} state re-isomerize thermally back to the ground
\textit{trans} state with some characteristic time. Upon removal of light, the electric polarization eventually
regains its initial dark state value.
Recently, it has been argued that the mechanism and details of light absorption are crucial to
understanding the optical depression of electric polarization.~\cite{knezevic:12}
In particular, it has been shown that the spontaneous polarization displays a non-uniform profile through
the sample.

\begin{figure}
  \includegraphics[width=6.6cm]{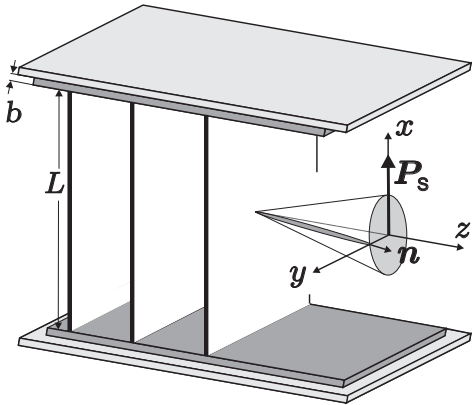}
  \caption{Bookshelf geometry of a SmC* sample with layers normal to the electrodes. The liquid
  crystal situated between glass plates, coated with insulating layers, is placed between two
  electrodes. The spontaneous polarization $\bm{P}\s{s}$ lies in the $xy$--plane, that is, in the
  plane of the smectic layers. To ensure that the polarization is directed along the $x$ axis,
  a voltage across the electrodes higher than $V\s{c} = 2b P\s{s}^0/(\vep_0 \vep_1)$ is needed.}
  \label{fig4}
\end{figure}

Here we suppose that the sample is confined between two parallel plates so that the smectic
layers are perpendicular to the plates -- the bookshelf geometry (Fig.~\ref{fig4}). The smectic layers usually shrink to some
extent at the tilting transition from the paraelectric SmA* to polar SmC* phase and form a folded chevron
structure.~\cite{rieker:87} Given that materials without substantial layer shrinkage~\cite{devries:79}
have also been found, in our simple analysis we shall neglect this shrinkage.
Charge pumps made of samples with large electric polarization in the direction perpendicular to
the electrodes have larger efficiency. A perpendicular polarization can be achieved by applying a voltage
between the electrodes. Indeed, a simple extension of the approach of reference [\onlinecite{copic:02}] reveals
that for voltages higher than $V\s{c} = 2 b P\s{s}^0 / (\vep_0 \vep_1)$ the sample's
polarization is perpendicular to the electrodes. For $b=10 \, {\rm nm}$, $\vep_1 = 10$,
and for a SmC* material with $P\s{s}^0 = 550 \, {\rm nC}/{\rm cm}^{2}$ we get a small voltage $V_c = 1.25 \, {\rm V}$.
We see that the polarization will have the required direction provided that the voltage $V_1$
(Fig.~\ref{fig2}) is greater than $V\s{c}$.

Typically one deals with SmC* hosts doped with dye molecules, whose number fraction
is a few percent. The behavior of the average polarization after time $t\s{on}$ can be described
theoretically.~\cite{knezevic:12} Here we shall simply suppose that the light is sufficiently
intense so that there is a time $t\s{on}$($<t\s{off}$) such that $\overline{P\s{s}}(t\s{on})$ vanishes.
In other words, we can regard $t\s{on}$ as the minimal time needed for a sufficient depletion of the \textit{trans}
dye population to eliminate the polarization.
Taking $\overline{P\s{s}}(t\s{on})=0$, formula (\ref{vonc}) for the voltage $V\s{on}$ reduces to
$V\s{on} = V_1 + LP\s{s}^0/\vep_0 \vep$.
For instance for a SmC* material with $P\s{s}^0 = 550 \, {\rm nC}/{\rm cm}^{2}$ and $\vep = 4$ in a cell of
thickness $L = 2 \, \mu {\rm m}$ one obtains a $V\s{on}-V_1 \approx 310 \, {\rm V}$, which is much larger
than $V_1$. Therefore, the optimal voltage $V_2$ is $V_2\sp{opt} \approx 155 \, {\rm V}$.

Supposing that $\overline{P\s{s}}(t\s{on})=0$, the efficiency (\ref{eta}) is
\be
\eta = \frac{(P\s{s}^0)^2}{4\vep_0 \vep U\s{in}}.
\label{etak}
\ee
Provided that $t\s{on}$ is shorter than the characteristic time of thermal
recovery of spontaneous polarization, and the process of back photoconversion from the \textit{cis} to the
\textit{trans} state of dye molecules can be neglected, the input energy can be expressed as $U\s{in} = n\s{d} \hbar \omega$.
Here $n\s{d}$ is the number of \textit{trans} dye molecules per unit volume that need to be
converted to \textit{cis} to eliminate $\overline{P\s{s}}$, and $\hbar \omega$ is the energy each
dye molecule absorbs from the light beam.

To get a numerical estimate of the efficiency we take as an example a polar SmC* host W314
doped with a few mol$\%$ of a racemic mixture of W470 azo dyes~\cite{walba:91,lanham:99, *chang:00}
(higher dye loadings lead to lower conversion due to flux attenuation~\cite{giesselmann:04,knezevic:12}). These two compounds have a
very similar structure, the N=N double bond of W470 being the only difference. The polarization
of the W314 host at $T=31 \, {}^{\circ} {\rm C}$ is roughly $550 \, {\rm nC}/{\rm cm}^2$ and is one of the highest outside bent-core
molecules.~\cite{niori:96} The \textit{trans} absorption maximum of W470 dyes is located at $365 \, {\rm nm}$,
which corresponds to photons of $3.4 \, {\rm eV}$. We assume that the quantum efficiency for the
\textit{trans} to \textit{cis} transition is equal to $1$. The relative molecular mass of W314
is $603$, and assuming a density of roughly $10^3 \, {\rm kg}/{\rm m}^3$, we obtain
a total molecular number density $n\s{tot} = 10^{27} \, {\rm molecules}/{\rm m}^3$.
If only, say, $2\%$ of molecules need to be converted to \textit{cis} to eliminate the polarization,
we obtain $n\s{d} = 2 \times 10^{25} \, {\rm molecules}/{\rm m}^3$. Taking the dielectric constant of W314 in the direction
perpendicular to the director $\bm{n}$ (see Fig.~\ref{fig4}) as $\vep = 4$, the efficiency (\ref{etak}) is $\eta \approx 2\%$.
Multiple layers with different dyes which absorb photons of different energies would
capture more of the broad solar spectrum.
A gradient of dye species can be stabilized by bonding them into an elastomeric SmC* host
-- a loosely linked solid that is still able to change its degree of order.
Such solids also reduce the risk of charge transport under fields ($V_2/L$ here) as in
dielectric actuation using rubber.~\cite{mckay:10}
Lower energy photons would give correspondingly higher efficiencies $\eta \sim 2-5\%$.

Let us briefly consider the case of a non-zero polarization profile through the sample.
We adopt a simple model for this profile, namely, we assume
that the polarization is eliminated in a layer of thickness $\Delta L$, while in the rest of the sample it is uniform
and equal to the dark state polarization $P\s{s}^0$. Then we can write
$\overline{P\s{s}} = \int_0^L P\s{s}(x) {\rm d} x/L = P\s{s}^0(L-\Delta L)/L$. As
$I\s{sun} t\s{on} A$ can be roughly estimated as $U\s{in} A \Delta L$,
for the power efficiency we get $\eta = (P\s{s}^0)^2 \Delta L/(4 \vep_0 \vep U\s{in} L)$.
The efficiency is, therefore, reduced by a factor of $\Delta L/L$ with respect to (\ref{etak}).

Our pump could equally work on cycles of temperature change that change the spontaneous
polarization. In this case the input energy is $U\s{in} = n\s{tot} \int C(T) {\rm d} T$, where
$C(T)$ is the specific heat per molecule at temperature $T$, and the integral goes from the dark state
temperature of ferroelectric material to the temperature where the spontaneous polarization is
eliminated (SmC*-SmA* phase transition temperature for liquid crystals).
The heat that is unavoidably generated by optical absorption thus adds to the efficiency
of our optical energy harvester.

In summary, we have proposed a light-powered charge pump for energy conversion, using a photoferroelectric
working material. The efficiency is explicitly given, which makes clear directions for further increase.

M.K. thanks the Winton Programme for the Physics of Sustainability and the Cambridge Overseas Trust for financial support, and M.W. thanks the Engineering and Physical Sciences Research Council (UK) for a Senior Fellowship. The authors thank
David Walba for advice.

\bibliography{Knezevic}
\end{document}